\documentclass[cameraready]{Interspeech}

\title{UniVoice: Unifying Autoregressive ASR and Flow-Matching based TTS with Large Language Models}

\author[affiliation={1,2}]{Wenhao}{Guan}
\author[affiliation={3,2}]{Zhikang}{Niu}
\author[affiliation={4}]{Ziyue}{Jiang}
\author[affiliation={1}]{Kaidi}{Wang}
\author[affiliation={1}]{Peijie}{Chen}
\author[affiliation={1}]{\\Qingyang}{Hong$^\ast$}
\author[affiliation={1}]{Lin}{Li$^\ast$}
\author[affiliation={3,2}]{Xie}{Chen}

\address{
  $^1$Xiamen University, China ~
  $^2$Shanghai Innovation Institute, China \\
  $^3$Shanghai Jiao Tong University, China ~ 
  $^4$Zhejiang University, China}
  \email{\{lilin,qyhong\}@xmu.edu.cn} 


\begin{document}

\maketitle

\begin{abstract}
    Large Language Models (LLMs) have become dominant in both Automatic Speech Recognition (ASR) and Text-to-Speech (TTS), but existing frameworks treat these tasks as isolated, overlooking inherent synergies. While discrete tokenization enables joint modeling, information loss compromises accuracy and synthesis quality. To address the limitations, we propose UniVoice, a unified LLM integrating ASR and TTS via continuous representations. Beyond discrete tokens, UniVoice preserves acoustic details for robust autoregressive recognition and high-fidelity flow-matching generation. To bridge architectural divergence between understanding and generation, we introduce a dual-attention masking mechanism to dynamically switch between causal masks for ASR and bidirectional masks for TTS. Additionally, text-prefix guided speech infilling ensures superior zero-shot voice cloning. Experimental results show UniVoice achieves competitive or superior performance to state-of-the-art single-task models, marking a significant step toward end-to-end speech interaction systems. The code and model weights are available at \url{https://github.com/gwh22/UniVoice}.

\end{abstract}

\section{Introduction}


In the realm of human-computer interaction, speech technologies play a pivotal role, with Automatic Speech Recognition (ASR) and Text-to-Speech (TTS) serving as two cornerstone components. ASR enables machines to understand spoken language, while TTS allows machines to generate natural-sounding speech. The development of unified architectures that can handle both ASR and TTS within a single framework is essential for advancing spoken language technologies. Such unification is particularly important for embodied intelligent human-computer interaction, where seamless and context-aware speech processing is critical.

Recent advances in Large Language Models (LLMs) have shown great potential for both ASR and TTS. For ASR, some models \cite{bai2024seed, tang2023salmonn} have demonstrated the effectiveness of LLMs in transcribing speech with high accuracy. 
SALMONN \cite{tang2023salmonn} uses a dual encoder from Whisper speech encoder \cite{radford2023robust} and BEATS audio encoder \cite{chen2022beats} for speech and  audio understanding. 
SeedASR is developed under the audio conditional LLM framework, utilizing the capabilities of LLM by submitting continuous speech representations, instructions, and contextual information into LLM.
In the realm of TTS, some frameworks \cite{jiang2023mega,wang2023neural,wang2024maskgct,du2024cosyvoice, anastassiou2024seed} have successfully adapted LLMs to generate high-quality speech from text. 
VALL-E \cite{wang2023neural} frames TTS as a conditional audio codec language modeling task with in context learning capabilities. 
CosyVoice \cite{du2024cosyvoice} proposes supervised semantic tokens for audio codec modeling and flow matching for acoustic details modeling.
These successes highlight the viability of leveraging LLMs for both ASR and TTS tasks.

Some works have begun to explore the use of language models as the backbone of the unified model for processing speech and text tasks \cite{ao2021speecht5,du2023lauragpt,tian2025opuslm,rubenstein2023audiopalm,maiti2024voxtlm,toyin2024sttatts}, thus completing various related tasks in the speech modality.
SpeechNet \cite{chen2021speechnet} and SpeechT5 \cite{ao2021speecht5} perform various speech tasks with an encoder-decoder model, specifically SpeechT5 needs to pretrain first and then finetune on subsequent tasks.
Viola \cite{wang2024viola} follows the VALL-E paradigm and integrates speech recognition, machine translation, and speech synthesis into a unified codec language model. LauraGPT \cite{du2023lauragpt} encodes the input audio into continuous representations using an audio encoder and generates the output audio from discrete codec codes.
OpusLM \cite{tian2025opuslm} is designed to accept and generate multistream discrete tokens in both text and speech modalities with pre-trained LLM. 
Some omni-modal conversational models \cite{xu2025qwen3,ai2025ming} aim to achieve end-to-end understanding and generation across all modalities. However, their massive parameter counts pose significant challenges for deployment on edge devices in real-world scenarios.
Despite these strides, discrete representations inherently suffer from information loss during quantization, which can limit reconstruction fidelity. To transcend these boundaries, a parallel research trajectory has embraced diffusion and flow matching (FM) models \cite{ho2020denoising,lipman2022flow}. Some systems \cite{chen2024f5,shen2023naturalspeech,guan2024reflow,lee2024ditto,le2024voicebox} have set new benchmarks for TTS by modeling speech within continuous feature spaces. However, while these continuous models excel in generative prowess, they often lack the autoregressive reasoning capabilities essential for ASR.

In this work, we propose UniVoice, a unified framework that operates entirely in continuous signal space, bypassing the limitations of discretization. 
Inspired by \cite{zhou2024transfusion,zhao2024monoformer}, UniVoice integrates autoregressive (AR) modeling for ASR and FM for TTS within a single Transformer backbone, leveraging the sequential stability of AR alongside the generative fidelity of FM. 

\begin{figure*}[t]
\centering
\includegraphics[scale=0.33]{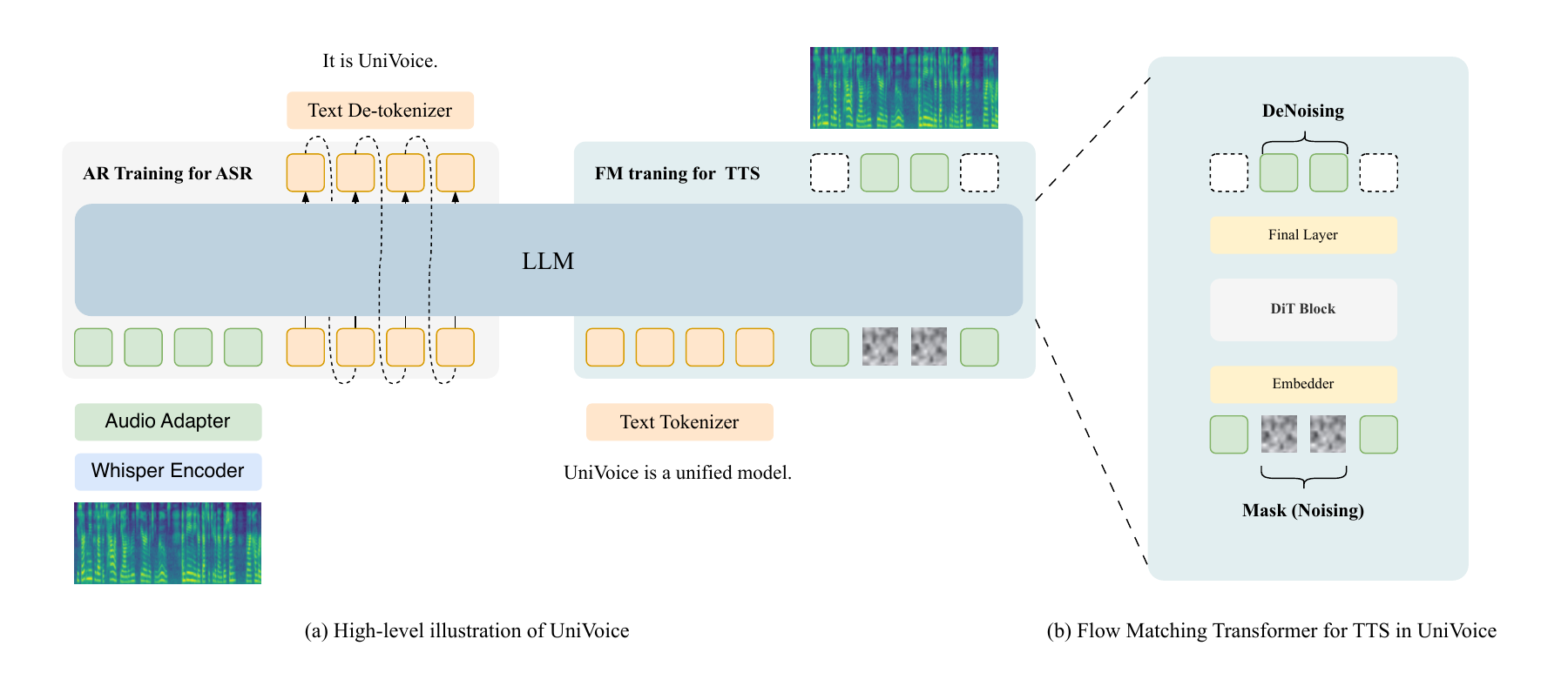} 
\caption{(a) An overview of UniVoice model. The Univoice model is composed of two branches: autoregressive (AR) Tranmsformer and flow matching (FM) transformer. The FM part is trained to denoise a masked noisy mel segment given its contextual counterpart, with prepending the text information. (b). The detailed
architecture of the flow matching transformer in the UniVoice model.}
\label{fig:overview}
\end{figure*}

The contributions of our work are as follows:

\begin{itemize}
    \item We present a novel architecture that integrates AR modeling for ASR with FM for TTS within a single pre-trained LLM. This allows UniVoice to operate in a continuous signal space, preserving fine-grained acoustic details and enabling joint speech understanding and generation.
    
    \item To bridge the gap between AR’s causal requirements and FM’s bidirectional nature, we propose a dual-attention mechanism that dynamically toggles masking strategies based on the task. Furthermore, we introduce text-prefix guided speech infilling, enabling robust, high-fidelity zero-shot TTS.
    
    \item Experiments trained on the LibriHeavy dataset \cite{kang2024libriheavy} demonstrate that our unified approach matches or surpasses specialized state-of-the-art models in both ASR and zero-shot TTS performance, proving that a single, parameter-efficient model can excel in both understanding and generative tasks.
\end{itemize}

\section{Model}

\subsection{Overview}
UniVoice is a dual-branch hybrid architecture designed to bridge speech understanding and generation. It comprises a causal Transformer for ASR and a flow-matching-based Diffusion Model for TTS, enabling the model to handle both tasks within a unified framework.

\subsection{Flow Matching}
The goal of Flow Matching is to learn a velocity field \(u_{t}^{\theta}\) parameterized by \(\theta\), which guides the evolution of samples over time, transforming samples from the source distribution \(X_{0} \sim p_{0}\) into samples from the target distribution \(X_{1} \sim q\).

The source distribution \(p_{0}\) is typically a standard normal distribution \(\mathcal{N}(x|0,1)\). The probability path \(p_{t}\) is constructed by aggregating conditional probability paths \(p_{t|1}(x|x_{1})\), each conditioned on a data example \(X_{1} = x_{1}\) from the training dataset. This path is also known as the conditional optimal transport path. Using this probability path, the random variable \(X_{t} \sim p_{t}\) is defined by drawing \(X_{0} \sim p_{0}\) and \(X_{1} \sim q\), and taking the linear combination \(X_{t} = tX_{1} + (1-t)X_{0}\).

The Flow Matching loss is defined as:
\begin{equation}
L_{FM}(\theta)=E_{t,X_{t}}||u_{t}^{\theta}(X_{t})-u_{t}(X_{t})||
\end{equation}
where \(u_{t}\) is the true velocity field that generates the desired probability path \(p_{t}\).

In practice, the true velocity field \(u_{t}\) is unknown. However, by conditioning on a single target example \(X_{1} = x_{1}\) from the training dataset, we can simplify the loss function. This leads to the conditional Flow Matching (CFM) loss:
\begin{equation}
L_{CFM}(\theta)=E_{t,X_{t},X_{1}}||u_{t}^{\theta}(X_{t})-u_{t}(X_{t}|X_{1})||^{2}
\end{equation}
Finally, using the conditional optimal transport path, we derive the Optimal Transport Conditional Flow Matching (OT-CFM) loss:
\begin{equation}
L_{CFM}^{OT}(\theta)=E_{t,X_{0},X_{1}}||u_{t}^{\theta}(X_{t})-(X_{1}-X_{0})||^{2}
\end{equation}
This loss function drives the learned velocity field to match the desired transformation from the source to the target distribution.

\subsection{Architectural Design}
The architecture of UniVoice is illustrated in Figure \ref{fig:overview}.

\subsubsection{Overall Architecture}
\begin{itemize}
    \item ASR Branch: This branch handles speech-to-text tasks by extracting acoustic features via a pre-trained Whisper encoder. These features are then processed by an adapter network, employing adaptive average pooling, to achieve temporal downsampling and semantic alignment with the LLM's latent space.
    
    \item TTS Branch (Text-prefix guided speech infilling): This branch reformulates speech synthesis as a flow-matching (FM) task, adopting an in-context modeling paradigm to maintain the structural integrity of the LLM. The input sequence follows a \textit{text-prefix strategy}, where transcript tokens precede speech features to serve as the textual conditioning signal. While our generative backbone is based on the Diffusion Transformer (DiT) \cite{Peebles_2023_ICCV}, we introduce critical modifications to ensure architectural homogeneity with the ASR branch. We omit the AdaLN-zero modulation mechanism characteristic of standard DiT. This design choice ensures that the Transformer blocks remain identical across both understanding and generation tasks, facilitating a truly unified backbone. To preserve the standard LLM input format, time embeddings are concatenated at the head of the noisy Mel-spectrogram sequence rather than being injected via global modulation. We integrate Rotary Positional Embeddings (RoPE) \cite{su2024roformer} within the self-attention layers to capture long-range dependencies. Furthermore, we implement a mask-and-infilling strategy inspired by Voicebox \cite{le2024voicebox,bai2022a3t} to predict speech features for masked segments, thereby significantly bolstering the model's capacity for high-fidelity, zero-shot voice cloning.
\end{itemize}

\begin{table*}[t]
    \centering
    \setlength{\tabcolsep}{2pt}  
    \renewcommand{\arraystretch}{1.1}  
    \scriptsize  
    \caption{Performance Comparisons of UniVoice and prior works. A-WER-clean, A-WER-other represents the ASR WER evaluation result on LibriSpeech test-clean dataset and LibriSpeech test-other dataset, respectively. }
    \begin{tabular}{c|c|cc|ccccc|cc}
        \toprule
        \textbf{Type} & \textbf{Method} & \textbf{Params} & \textbf{Data-hrs} & \textbf{SIM$\uparrow$} & \textbf{WER$\downarrow$} & \textbf{UTMOS$\uparrow$} & \textbf{SMOS$\uparrow$} &  \textbf{MOS$\uparrow$} & \textbf{A-WER-clean$\downarrow$} & \textbf{A-WER-other$\downarrow$} \\
        \midrule
        & Ground Truth & - & - & 0.69  & 2.43 & 4.07 & 3.82 & 4.33$\pm$0.10 & - & - \\
        \midrule
        \multirow{5}{*}{{Unified Models}}
        & SpeechT5 & $0.14$B & 0.96K & 0.33 & 5.91 & 3.32 & 3.35 & 3.36$\pm$0.09 & 4.4 & 10.4 \\
        & LauraGPT & $2.0$B & 60K & - & 8.62 & - &  - & - & 4.4 & 7.7 \\
        & OpusLM-0.4B & $0.4$B & 213K & - & 19.8 & - & - & - & 4.2 & 8.7 \\
        & OpusLM-7B & $7$B & 213K & - & 4.60 & - &  - & - & \textbf{2.3} & \textbf{5.2} \\

        & UniVoice (Ours) & 0.4B & 50K & \textbf{0.56} & \textbf{4.06} & \textbf{3.72} & \textbf{3.88} & \textbf{3.96$\pm$0.06} &  \underline{3.0} & \underline{6.3} \\

        \midrule
        \midrule
        \multirow{8}{*}{{Only TTS Models}} 
        & CosyVoice & $0.4$B & 170K & 0.66 & 3.59 & 4.17  & \underline{3.96} & \underline{4.15$\pm$0.07} & - & - \\
        & CosyVoice2  & $0.6$B & 167K & \textbf{0.66} & \textbf{2.23} & \textbf{4.38} & \textbf{3.98} & \textbf{4.29$\pm$0.09} & - & -  \\
        & MaskGCT & $1.0$B & 100K & 0.66 & \underline{2.49} & 3.85  & 3.92 & 4.12$\pm$0.06 & - & - \\
        & F5-TTS & $0.3$B & 100K & 0.66 & 2.54 & 3.84  & 3.90 &  3.98$\pm$0.10 & - & -  \\
        & FireRedTTS & $0.6$B & 248K & 0.47 & 2.69 & 3.91  & 3.85 & 4.01$\pm$0.09 & - & -  \\
        & Spark-TTS & $0.3$B  & 102K & 0.55 & 2.89 & \underline{4.23} & 3.95 & 4.11$\pm$0.08  & - & -  \\

        & UniVoice-TTS~(Ours)& $0.4$B & 50K & 0.56 & 4.66 & 3.92  & 3.86 & 4.06$\pm$0.06 & - & -  \\
        \midrule
        \midrule
        \multirow{8}{*}{{Only ASR Models}} 
        & Whisper-small & $0.2$B & 680K & - & - & - & - &  - & 3.4 & 7.6  \\
        & Whisper-large-v2 & $1.5$B & 680K & - & - & - & - &  - & 2.7 & 5.2 \\
        & Whisper-large-v3 & $1.5$B & 680K & - & -  & - & - &  - & 1.9 & \underline{3.6} \\
        & Whisper-large-v3-turbo & $0.8$B & 680K & - & - &  - &  - & - & \textbf{1.9} & \textbf{3.5} \\
        & Paraformer & $0.2$B & 20K & - & - & - & - &  - & 3.5 & 8.2 \\
        & Zipformer & $0.15$B & 0.96K & - & - & - & - &  - & \underline{2.0} & 4.4 \\

        & UniVoice-ASR~(Ours) & $0.4$B & 50K & - & - &  - & - & - & 2.5 & 4.2  \\

        \bottomrule
    \end{tabular}
    \label{tab:exp-geneval}
\end{table*}

\subsubsection{Dual-Attention Masking}

To facilitate the simultaneous optimization of speech understanding and generation, we meticulously design the attention masking strategy to accommodate the distinct requirements of each task. For ASR, UniVoice employs a standard causal mask, consistent with the autoregressive nature of LLMs. Conversely, for TTS, we implement a bidirectional attention mask to fully capture the global context of the text. Inspired by the in-context learning paradigm in VALL-E \cite{wang2023neural}, this approach leverages the LLM's inherent text-modeling strengths to condition the speech synthesis process. The synergy of these task-specific strategies constitutes our proposed Dual-Attention Masking mechanism.

\subsection{Training Strategy}
The model is trained using a multi-task objective to effectively integrate audio and text modalities.

\begin{itemize}
    \item ASR Training: Employs an autoregressive objective where a standard Transformer, enhanced by an audio encoder and adapter, predicts text token probabilities from audio embeddings. We define the ASR loss as \(L_{LM}(\theta)\), which follows the training strategy of original LM.
    \item TTS Training: Uses the OT-CFM framework to frame speech generation as a text-prefix guided infilling task. The model learns to reconstruct masked speech \(m \odot X_{1}\) conditioned on both the unmasked context \(X_{ctx}\) and text tokens \(z\).
\end{itemize}

The unified loss function is defined as a weighted sum:
\begin{equation}
L_{total} = \lambda L_{LM}(\theta) + L_{audio}^{cfm}(\theta)
\end{equation}
where \(L_{LM}\) is the autoregressive loss for ASR, and \(L_{audio}^{cfm}\) is the flow matching loss:
\begin{equation}
L_{audio}^{cfm}(\theta) \\ 
= E_{t,X_{0},X_{1},m}||m \odot (u_{t}^{\theta}(X_{t},X_{ctx},z)-(X_{1}-X_{0}))||^{2}
\end{equation}

Classifier-Free Guidance \cite{ho2022classifier} (CFG) is incorporated by randomly dropping conditions during training to enhance generation quality.

\subsection{Inference Process}
During inference, the model operates in two modes depending on the task.

\begin{itemize}
    \item ASR Inference: Performs conventional autoregressive decoding, iteratively sampling the most probable tokens conditioned on the input audio features.
    \item TTS Inference: Follows a prompt-based infilling workflow. Given a reference audio (\(X_{ref}\)), its transcript (\(Y_{ref}\)), and the target text (\(Y_{gen}\)), the model calculates the target duration based on the ratio \(\text{ratio} = \frac{\text{len}(Y_{gen})}{\text{len}(Y_{ref})}\). It constructs the text condition \(z\) and acoustic context \(X_{ctx}\), solves the flow ODE starting from Gaussian noise \(X_0\) to reach the target mel-spectrogram \(X_1\), and converts the resulting mel-spectrogram into a waveform using a neural vocoder.
\end{itemize}

By integrating these components, UniVoice effectively bridges the gap between speech recognition and generation, providing a robust framework for handling complex tasks.

\section{Experiments}

\subsection{Experimental Setup}
We trained UniVoice on the 50K-hour LibriHeavy dataset. For TTS, the original 16kHz audio was upsampled to 22.05kHz, and for ASR, it was maintained at 16kHz. Evaluations were conducted using LibriSpeech-PC test set for zero-shot TTS and the LibriSpeech \cite{panayotov2015librispeech} test-clean/test-other subsets for ASR. All audio was processed into 80-bin mel-spectrograms with a 1024 frame size and a 256 hop size.

\subsubsection{Implementation Details}
UniVoice uses SmolLM2-360M \cite{allal2025smollm2smolgoesbig} as the backbone language model, paired with a Whisper-large-v3-turbo encoder for audio feature extraction and BigvGAN \cite{lee2022bigvgan} for waveform reconstruction. The model was trained for 10 epochs using the AdamW optimizer with a learning rate of 1.5e-3, \(\beta_{1}=0.9\), \(\beta_{2}=0.95\), and a 20,000-step warmup, alongside a cosine scheduler. The batch size was 160,000 audio frames. ASR was assigned a secondary task weight of \(\lambda=0.005\), while TTS training involved masking 70\% to 100\% of Mel-spectrogram frames. During inference, CFG weight was applied with probabilities of 0.2 for text and 0.3 for masked speech tokens, using a CFG weight of 2 and 32 sampling steps.

For baseline comparison, UniVoice-TTS-speaker was designed with a simplified architecture. It generates mel-spectrograms directly from text embeddings, using speaker embeddings derived from the first layer of the XLSR-53 model \cite{conneau2020unsupervised} to capture speaker timbre, thus supporting voice cloning. This approach bypasses masked infilling paradigms for a more straightforward mel-spectrogram generation process.

\subsubsection{Evaluation Metrics}
Zero-shot TTS was evaluated using UTMOS and MOS for naturalness, cosine similarity (SIM) and SMOS for speaker similarity, and WER for robustness. ASR performance was measured by WER on LibriSpeech test-clean/test-other, using Whisper-large-v3 for transcription.

\subsubsection{Baselines}
UniVoice was compared to several baselines: unified models (SpeechT5 \cite{ao2021speecht5}, LauraGPT \cite{du2023lauragpt}, OpusLM \cite{tian2025opuslm}), zero-shot TTS models (F5-TTS \cite{chen2024f5}, CosyVoice \cite{du2024cosyvoice}, CosyVoice2 \cite{du2024cosyvoice2}, SparkTTS \cite{wang2025spark}, FireRedTTS \cite{guo2024fireredtts}, MaskGCT \cite{wang2024maskgct}), and ASR models (Whisper series, Paraformer \cite{gao2022paraformer}, Zipformer \cite{yao2023zipformer}). UniVoice-TTS and UniVoice-ASR were trained solely for TTS and ASR within our framework, which were also included as baselines.

\subsection{Main Results}

\begin{table}[t]
\centering
\setlength{\tabcolsep}{2pt}  
\renewcommand{\arraystretch}{1.1}  
\scriptsize  
\caption{Comparison of the zero-shot TTS capability of different TTS model variants of UniVoice on LibriSpeech-PC test set.}
\begin{tabular}{lccc}
\toprule
\textbf{Method} & \textbf{WER\(\downarrow\)} & \textbf{SIM\(\uparrow\)} & \textbf{UTMOS\(\uparrow\)} \\ 
\midrule
UniVoice-TTS-speaker  & 5.72  & 0.29 & 3.65 \\
UniVoice-TTS-infilling  & \underline{4.66}  & 0.56 & \textbf{3.92} \\
UniVoice  & \textbf{4.06}  & \textbf{0.56} & \underline{3.72} \\
\bottomrule 
\end{tabular}

\label{tab:tts_comp}
\end{table}

\begin{table}[t]
\centering
\setlength{\tabcolsep}{2pt}  
\renewcommand{\arraystretch}{1.1}  
\scriptsize  
\caption{Comparison of performance using different attention masks for TTS on LibriSpeech-PC test set.}
\begin{tabular}{lccc}
\toprule
\small
\textbf{Method} & \textbf{WER$\downarrow$} & \textbf{SIM$\uparrow$} & \textbf{UTMOS$\uparrow$} \\ 
\midrule
AR Mask  & 9.85  & 0.49 & 2.23  \\
Full Mask  & \textbf{4.66}  & \textbf{0.56} & \textbf{3.92}   \\
\bottomrule 
\end{tabular}

\label{tab:attn_mask}
\end{table}

\begin{table}[t]
\centering
\setlength{\tabcolsep}{2pt}  
\renewcommand{\arraystretch}{1.1}  
\scriptsize  
\caption{Comparison of performance using different \(\lambda\). WER-c and WER-o represent the ASR WER evaluated on LibriSpeech test-clean dataset and test-other dataset, respectively.}
\begin{tabular}{lccc|cc}
\toprule
\textbf{Method} & \textbf{WER$\downarrow$} & \textbf{SIM$\uparrow$} & \textbf{UTMOS$\uparrow$} & \textbf{WER-c$\downarrow$} & \textbf{WER-o$\downarrow$} \\ 
\midrule
\(\lambda=0.01\)  & 4.66  & 0.54 & 3.69 & 4.21 & 7.82 \\
\(\lambda=0.005\)  & \textbf{4.06}  & \textbf{0.56} & \textbf{3.72} & \textbf{3.01} & \textbf{6.36}  \\
\bottomrule 
\end{tabular}

\label{tab:lambda}
\end{table}

\subsubsection{Zero-Shot TTS Results}
As shown in Table \ref{tab:exp-geneval}, UniVoice demonstrated strong robustness with a 12\% relative WER reduction on LibriSpeech test-clean compared to advanced unified models. It also outperformed UniVoice-TTS by 13\% in WER, indicating that joint ASR-TTS training improves intelligibility through shared linguistic representations. Although there is a slight performance gap compared to specialized single-task systems, likely due to parameter sharing and task competition, UniVoice effectively denoises during mel-spectrogram synthesis.

In terms of speaker similarity, UniVoice achieved competitive results but lagged behind some baselines like CosyVoice and F5-TTS. This might stem from the absence of the AdaLN-zero modulation mechanism, which is crucial for speaker-specific adaptation.
For speech quality, UniVoice achieved the highest UTMOS among unified models but showed a 0.2 UTMOS decrease compared to UniVoice-TTS. This suggests that joint training requires a trade-off in naturalness.

\subsubsection{ASR Results}
UniVoice was compared with state-of-the-art models, including Whisper-small, Whisper-large-v2, Whisper-large-v3, Whisper-large-v3-turbo, Paraformer, and Zipformer, as well as previous unified models. The results indicate that UniVoice achieves excellent audio comprehension despite being a relatively small model trained on a smaller dataset. As shown in Table \ref{tab:exp-geneval}, compared to UniVoice-ASR, UniVoice performs slightly worse due to the competing objectives of joint training.

\subsubsection{Ablation Study}
We conducted ablation studies to validate the design choices of our proposed UniVoice model. 

\textbf{Ablation Study of TTS model variants}
As shown in Table \ref{tab:tts_comp}, we compared two variants of UniVoice: a speech-infilling model and a speaker-embedding-conditioned baseline. Results showed that the infilling approach significantly outperformed the baseline across all evaluation metrics. It achieved an 18\% reduction in WER and improvements in SIM and UTMOS by 0.27. This demonstrates the effectiveness of the speech-infilling paradigm and its integration within our unified architecture.

\textbf{Ablation Study on Attention Mask Strategies}
We explored different attention mask configurations for our text-prefix-guided speech-infilling TTS approach. As shown in Table \ref{tab:attn_mask}, our findings confirmed that using full bidirectional attention masks consistently outperformed autoregressive masking in all metrics, including WER, SIM, and UTMOS. This result underscores the importance of complete context access for achieving high-quality speech synthesis.

\textbf{Ablation Study on Different \(\lambda\)}
We evaluated the impact of the balancing parameter $\lambda$ on the joint optimization of ASR and TTS objectives.
As shown in Table \ref{tab:lambda}, experiments showed that setting \(\lambda\) to 0.005 yielded better performance than 0.01. We attribute this to the higher complexity of training TTS based on flow matching, which benefits from a larger weight. In contrast, ASR tasks are relatively simpler and thus assigned a smaller weight. This arrangement allows the model to prioritize learning the more challenging TTS task during training.

These ablation studies collectively highlight the importance of our design choices, including the speech-infilling mechanism, full bidirectional attention masks, and careful tuning of task weighting. They provide insights into optimizing our model for improved performance in both TTS and ASR tasks.

\section{Conclusion}
This paper introduces UniVoice, a unified Transformer framework integrating autoregressive ASR with flow-matching-based TTS. By employing a dual attention mechanism and text-prefix-guided speech-infilling, UniVoice effectively balances causal and bidirectional patterns for high-fidelity speech understanding and zero-shot synthesis. Our results validate the synergy between these complementary paradigms in a single LLM-based architecture.
Despite its performance, UniVoice currently focuses on ASR and TTS. Future work will extend this framework to a broader speech processing tasks. 
To support reproducibility, we will open-source our code and checkpoints. Demo samples are available at: \url{https://univoice-demo.github.io/UniVoice }.

\section{Generative AI Use Disclosure}
During the preparation of this manuscript, the authors used generative AI tools to polish the English language, improve readability, and assist with LATEX formatting. These tools were not used to generate any scientific claims, experimental results, or significant parts of the manuscript.

\bibliographystyle{IEEEtran}
\bibliography{mybib}

\end{document}